\documentclass[mathleft,fleqn,%
]{an}
\usepackage{amsmath,amssymb,url}
\usepackage{graphicx,color}
\usepackage{natbib}
\usepackage[varg]{txfonts}

\newcommand{\XMM}{\emph{XMM-Newton}}
\begin{document}

% The following seven commands are intended for editorial usage and
% should be ignored by the author(s).
\Pagespan{1}{}% Document's page range. 
% If second parameter is left empty, the last page is computed
% automatically.
\Yearpublication{2020}%
\Yearsubmission{2020}%
\Month{0}%   
\Volume{999}%
\Issue{0}%  
\DOI{asna.201900000}% 

\title{MACHO 311.37557.169: A VY~Scl star}

\author{H. W\"orpel\inst{1}\fnmsep\thanks{Corresponding author:
        {hworpel@aip.de}}
\and A. D. Schwope\inst{1}
\and I. Traulsen\inst{1}
\and M. J. I. Brown\inst{2}
}
\titlerunning{MACHO 311}
\authorrunning{H. Worpel, A. Schwope, I. Traulsen, M. Brown}
\institute{
Leibniz-Institut f\"ur Astrophysik Potsdam (AIP), An der Sternwarte 16, Potsdam, 14482, Germany \and 
School of Physics \& Astronomy, Monash University, Clayton, Victoria 3800, Australia}
\received{}
\accepted{}
\publonline{}

\date{accepted}
\abstract{Optical surveys, such as the MACHO project, often uncover variable stars whose classification requires followup observations by other instruments. We performed X-ray spectroscopy and photometry of the unusual variable star MACHO 311.37557.169 with \XMM\ in April 2018, supplemented by archival X-ray and optical spectrographic data. The star has a bolometric X-ray luminosity of about $1\times 10^{32}$\,erg\,s$^{-1}$\,cm$^{-2}$ and a heavily absorbed two-temperature plasma spectrum. The shape of its light curve, its overall brightness, its X-ray spectrum, and the emission lines in its optical spectrum suggest that it is most likely a VY~Scl cataclysmic variable.}

\maketitle

\section{Introduction}

The MACHO survey \citep{AlcockEtAl1992, AlcockEtAl1997, AlcockEtAl2000} was a two-colour photometric
study of several million stars in the Magellanic Clouds and the Galactic Bulge that aimed to spot
gravitational lensing events associated with massive free-floating bodies in the Galactic halo.
A useful byproduct was the discovery of numerous intrinsically variable stars in the southern sky (e.g. \citealt{CieslinskiEtAl2004}), of which many still require classification. 

Our target was first observed as a variable star of unknown nature by \cite{Hoffleit1972}, ranging between visual magnitudes 16.2 and 14.8 on photographic plates, and was eventually designated NSV~10530 (e.g. \citealt{SamusEtAl2004}). It was later observed during the MACHO survey, under the designation {MACHO~311.37557.169} (henceforth M311, {RA 18 18 41.7}, {DEC -23 56 21.10}), and was identified as a possible cataclysmic variable by \cite{ZaniewskiEtAl2005}. Those authors were searching for R~CrB stars, obtained optical spectra of numerous candidates, and
excluded this object because it exhibits the conspicuous Balmer lines that R~CrBs lack. Instead,
they hypothesised that it is an AM~Her star.  The Gaia Data Release 2 \citep{Gaia2016, Gaia2018,Bailer-JonesEtAl2018} gives a parallax of $0.723\pm 0.054$\,mas for M311, providing a distance estimate of $1.34^{+0.11}_{-0.09}$\,kpc. The star is listed as bright as $m_V=14.8$ in DR10 of the APASS catalogue \citep{HendenEtAl2018}, at $m_V=15.7$, $m_I=15.2$ in the OGLE-III database \citep{SzymanskiEtAl2011}, and at around $m_V=14.8$ in ASASSN \citep{ShappeeEtAL2014, KochanekEtAl2017}.

M311 has been observed by XMM-Newton on two occasions. In 2006 it was spotted serendipitously in an observation of the pulsar candidate AX~J1817.6$-$2401, in which M311 is visible far off-axis, about 11.7 arcminutes, with the MOS1 and MOS2 cameras and was thus designated as {3XMM~J181841.7-235618} in the 3XMM catalogue \citep{RosenEtAl2016}. It is unfortunately outside the fields of view of the EPIC-$pn$ and Optical Monitor. A second, targeted, pointing was performed in 2018 by XMM-Newton and we additionally found a short serendipitous observation of its field by Swift from Aug 2017.

We here present an analysis of the X-ray data, the MACHO light curve, and the optical spectrum. Our aim is to classify M311.

\section{Method and Results}

\subsection{Optical spectrum}

\begin{figure}
    \includegraphics{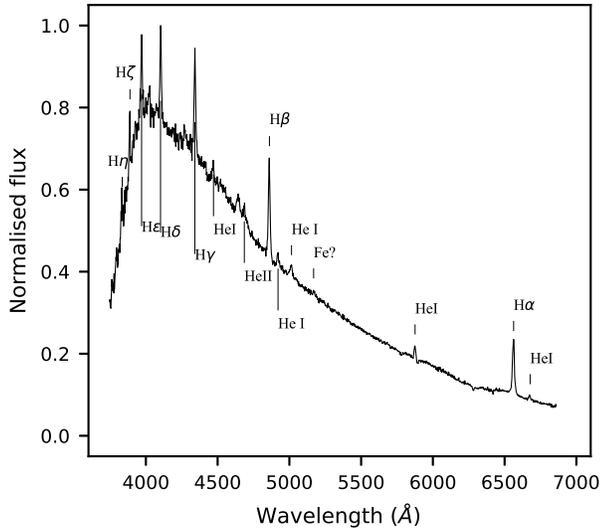}
    \caption{Optical spectrum of M311. The units of flux are arbitrary.}
    \label{fig:optspec}
\end{figure}

\cite{ZaniewskiEtAl2005} obtained an identification spectrum of M311 with the LDSS2 spectrograph of
the Magellan telescope at Las Campanas, which they generously made available to us. The 900\,s observation was performed on 2003 May 10. Unfortunately no standard star spectra are available for this observation so we are unable to perform a proper flux calibration. We were, however, able to get an adequate wavelength calibration by identifying hydrogen Balmer lines in the spectra by eye and fitting their known wavelengths to the CCD pixel values with a low order polynomial. The spectrum is shown in Figure \ref{fig:optspec}.

The hydrogen Balmer lines are very prominent and the helium lines, though clearly present, are quite faint. These features are reminiscent of a cataclysmic variable (CV). 

\subsection{Optical photometry}

\begin{figure}
    \includegraphics{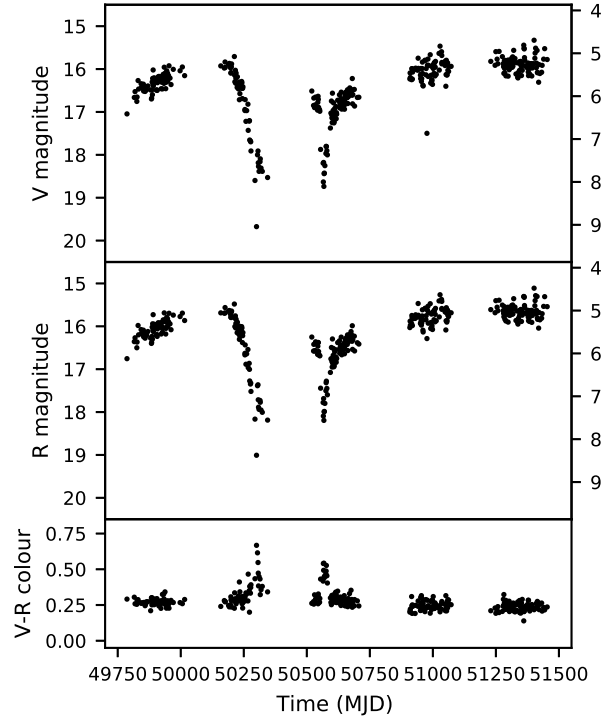}
    \caption{Optical light curves of M311 in converted $V$ and $R$ magnitudes. Apparent and absolute (for $d=1.34$\,kpc) magnitudes are indicated on the left and right axes respectively.}
    \label{fig:optlicu}
\end{figure}

We downloaded the light curves of M311 from the MACHO survey website and corrected
the timings to the solar system barycenter using the algorithms of \cite{EastmanEtAl2010}. The
MACHO blue and red filter magnitudes were converted to Johnson $V$ and Kron-Cousins
$R$ magnitudes using the conversion formulae in \cite{PopowskiEtAl2003}. The light curves and colours are shown in Figure \ref{fig:optlicu}. We also include absolute magnitudes, derived from the Gaia distance. These have not been corrected for Galactic extinction.

To find possible periodicities of a few hours in the long-term light curve, we subtracted the best-fitting quartics from the segments before and after the dip and performed an analysis-of-variance (AoV; e.g. \citealt{Schwarzenberg-Czerny1989}) on the residuals. Other than a signal at 24 hours and some of its integer divisors-- the rotation of the Earth and its aliases-- we found no signal.

By adding a faked sinusoidal signal to the residuals we deduced that a modulation with a half-amplitude of 0.1 magnitudes would have been detectable if stable in time.

The ASASSN survey has observed the target 290 times, during which it had an average $m_V$ of 14.8 and no long-term light curve variations. We downloaded these data and performed another AoV search on them between 5 minutes and 24 hours to search for shorter term periodicities, but there was no significant signal.

\subsubsection{Hutton-Westfold Observatory Data}

The Hutton-Westfold Observatory is a teaching observatory located at Monash University in Melbourne, Australia. It consists of a 14-inch telescope. On the night of 2019 Oct 23 we obtained a single 60\,s exposure of M311 that confirmed the target was in a bright state.

The following night we obtained $63\times 60$\,s exposures using the V filter. One exposure, near the end of the run, was affected by a passing cloud and unusable. We performed bias, dark, and flat field correction, and image alignment using AstroImageJ \citep{CollinsEtAl2017}. The observations were performed under very challenging conditions, with significant light pollution, some high cloud, and high airmass (1.46--2.01).

We used a nearby bright star at RA~18~18~42~DEC~-23~52~43 with V magnitude 9.798$\pm0.004$ \citep{HendenEtAl2018} as the comparison star. The mean magnitude, measured from the stacked observations, of M311 is around $m_V=15.2\pm 0.1$, slightly brighter than it was during the the MACHO observations. We used apertures of radius 40 and 19 pixels for the comparison star and target respectively, and annuli of outer radius 58 and 29 pixels respectively for the sky.

We did not see any clear evidence for variability of M311 in the individual frames, due to the unfavourable viewing conditions. Stacking multiple frames together did not help, so we are unable to detect or to rule out variability of $\sim 0.2$ magnitudes or less.

\subsection{X-ray spectra}

\emph{XMM-Newton} observed M311 for 13\,ks on 2006 March 19 (obsid 0304220401) and for 23\,ks on 2018 Apr 11 (obsid 0803830201). The Swift observation was 0.5\,ks long, on 2017 Aug 29 (obsid 07014478001). The observations are summarised in Table \ref{tab:obslog}.

\begin{table*}
   \caption{X-ray observation log for M311}
   \begin{tabular}{lllll}
   Date        & OBSID       & Inst  & Camera/Filter/Mode           & Exposure (ks) \\\hline
   2006 Mar 19 & 0304220401  & XMM-Newton   & EPIC-MOS1/medium/full frame  & 13.1          \\
               &             &       & EPIC-MOS2/medium/full frame  & 13.1          \\\hline
   2017 Aug 29 & 07014478001 & Swift & XRT/photon counting          &  0.5          \\\hline
   2018 Apr 11 & 0803830201  & XMM-Newton   & EPIC-MOS1/thin/full frame    & $14.5+1.7$    \\
               &             &       & EPIC-MOS2/thin/full frame    & 18.6          \\
               &             &       & EPIC-pn/thin/full frame      & 17.5          \\
               &             &       & OM/UVW1/fast                 & $4.4+2\times3.35$   \\
              
   \end{tabular}
   \label{tab:obslog}
\end{table*}

We reduced the \emph{XMM-Newton} data with version 16.1.0 of the XMM-SAS
software and produced photon event lists with the \emph{emchain} and \emph{epchain} tasks. The arrival times
were corrected to the solar system barycenter with the \emph{barycen} task.
In the earlier observation the source was so far off-axis that the point spread function was
distinctly non-circular, so we used an elliptical source extraction
region with minor and major radii (10 and 15 arcsec)
respectively, and rotated to approximately the same orientation as the source.
The background extraction region was a large circle located in a source-free region on the same
chip. For the later, targeted, observation we used circular source and background extraction regions.

\begin{figure*}
    \includegraphics{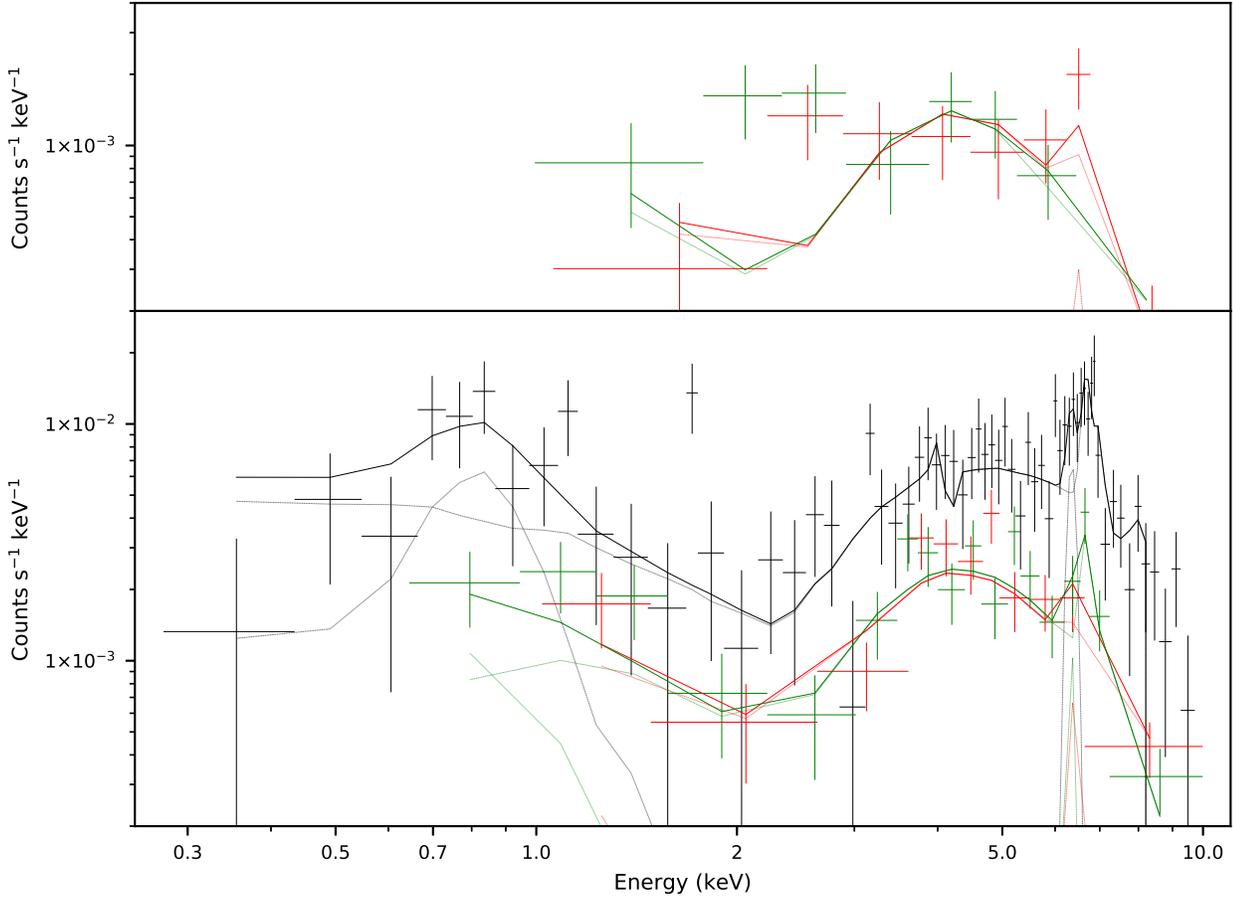}
    \caption{X-ray spectra of M311 for 2006 (top panel) and 2018 (bottom panel). The contributions of the individual additive components are indicated with thinner lines.}
    \label{fig:xrayspec}
\end{figure*}

The spectra are shown in Figure \ref{fig:xrayspec}. We assumed the source had the same spectrum, with possibly varying intensity, in both observations so we fitted all five spectra jointly. We attempted to fit with a Mekal plasma model \citep{MeweEtAl1985,LiedahlEtAl1995} and found that a two-temperature plasma was necessary together with strong partial absorption. We also required an additional Gaussian near 6.4\,keV. Thus, the Xspec model was {\tt const*pcfabs*(mekal+mekal+gaussian)}.

The results are given in Table \ref{tab:specfits}. Uncertainties are at the $1\sigma$ level and fluxes are bolometric, obtained via the \texttt{cflux} command of {\sc Xspec} \citep{Arnaud1996}. The equivalent width of the additional Gaussian component was determined using the \texttt{eqwidth} command and found to be $350\pm100$\,eV.

The variable normalisation factor between the 2018 and 2006 observation was $1.14^{+0.13}_{-0.12}$, indicating that the 2006 observation may have been slightly brighter, but the results are consistent with a constant X-ray luminosity. The absorption fraction, though very close to unity, did not give an adequate fit if we set it to 100\%, or replaced it with a totally covering cold absorber. We also calculated the X-ray luminosity of M311 using the bolometric flux and the Gaia distance.

\begin{table}
\caption{Spectral fit for M311. Fluxes are bolometric, and uncertainties are at the $1\sigma$ level.}
\begin{tabular}{lc}\vspace{1mm}
$N_\textrm{H}$ & $15.5^{+1.3}_{-1.1}\times 10^{22}$\,cm$^{-2}$\\\vspace{1mm}
Covering Fraction & $98.34^{+0.43}_{-0.44}$\% \\\vspace{1mm}
Plasma Temp 1 & $0.59^{+0.23}_{-0.11}$\,keV\\\vspace{1mm}
Plasma Norm 1 & $1.15\pm0.34\times 10^{-4}$ \\\vspace{1mm}
Plasma Temp 2 & $7.8^{+0.9}_{-1.0}$\,keV\\\vspace{1mm}
Plasma Norm 2 & $9.8^{+0.4}_{-0.5}\times 10^{-4}$ \\\vspace{1mm}
Line Energy & $6.36^{+0.06}_{-0.07}$\,keV\\\vspace{1mm}
Line norm   & $4.2^{+1.2}_{-1.1} \times 10^{-6}$ \\\vspace{1mm}
Line eq. wd. & 0.35$\pm$0.1 \,keV \\\vspace{1mm}
$\chi^2_\nu$ & 1.167 (100) \\\vspace{1mm}
Absorbed flux & $1.13\pm 0.04 \times 10^{-12}$\,erg\,s$^{-1}$\,cm$^{-2}$ \\\vspace{1mm}
Unabsorbed flux & $3.45\pm0.12 \times 10^{-12}$\,erg\,s$^{-1}$\,cm$^{-2}$ \\\vspace{1mm}
Absorbed luminosity & $2.4^{+0.29}_{-0.25}\times 10^{32}$\,erg\,s$^{-1}$ \\\vspace{1mm}
Unabsorbed luminosity & $7.41^{+0.90}_{-0.75}\times 10^{32}$\,erg\,s$^{-1}$ \\\vspace{1mm}
\end{tabular}
\label{tab:specfits}
\end{table}

Simplifying the model, by changing to a one-temperature plasma or replacing the partially covering absorber with a cold totally-covering absorber, did not produce acceptable fits. We obtained $\chi^2_\nu$ of 1.84 and 1.30 respectively, and the cooler hump in the 2018 spectrum was clearly not fit adequately.

The binned spectra clearly show a feature around 6.0-7.0\,keV, coincident with the iron emission line triplet. If, in particular, the 6.4\,keV fluorescence line is present we can use its equivalent width to determine if the X-ray emission is primarily scattered, as is sometimes seen in CVs with discs. We therefore tested whether the iron lines are significantly detected, under the following assumptions:

First, we assume that the emission between 5.5\,keV and 7.5\,keV is an unabsorbed plasma continuum with possibly Gaussian emission lines superimposed on it. We model the plasma continuum as a bremsstrahlung with temperature fixed at 7.8\,keV. For this exercise we will not use a Mekal model since that already includes the 6.7 and 6.9\,keV lines. Second, we assume that the spectrum is the same shape in 2018 as it was in 2006 but with possibly different luminosity. Thus, we separated the 2018 and 2006 data into two spectral groups, with a constant multiplicative factor that can differ between them, as for the previous fit.

To avoid losing fine spectral features to the binning procedure, we fit the unbinned spectra between 5.5\,keV and 7.5\,keV with the $c$-statistic. We used the method developed by \cite{Kaastra2017} to estimate the goodness of fit. Fitting with only a bremsstrahlung model and no Gaussians gave a $c$-stat $4.1\sigma$ above the expected value, indicating a poor fit. Thus, the iron line complex as a whole is clearly detected.

Next, we added a single Gaussian to test the possibility that the iron line triplet is detected but that the individual lines cannot be distinguished. For a Gaussian with best-fit energy $6.68^{+0.16}_{-0.21}$\,keV and equivalent width of $1.4^{+1.6}_{-0.9}$\,keV-- for all three lines combined-- we obtained a fit consistent (0.45$\sigma$) with the data. We conclude therefore that there is no need to separate the iron line complex into three lines to obtain a formally acceptable fit and that, therefore, we cannot resolve the individual lines.

Furthermore, this fit gave a line energy of around 6.7\,keV-- suggesting that the fluorescent line is approximately equal in intensity to the 7.0\,keV line, or around 0.2\,keV and somewhat lower than the more crude fit performed above. These two lines have equivalent widths of 650 and 340\,eV respectively for a 7.8\,keV Mekal. Thus, the sum of the equivalent widths is consistent with the value derived above, with or without the 6.4\,keV fluorescent line. We conclude that evidence for its presence is weak at best.

\subsection{X-ray photometry}

\begin{figure}
\includegraphics{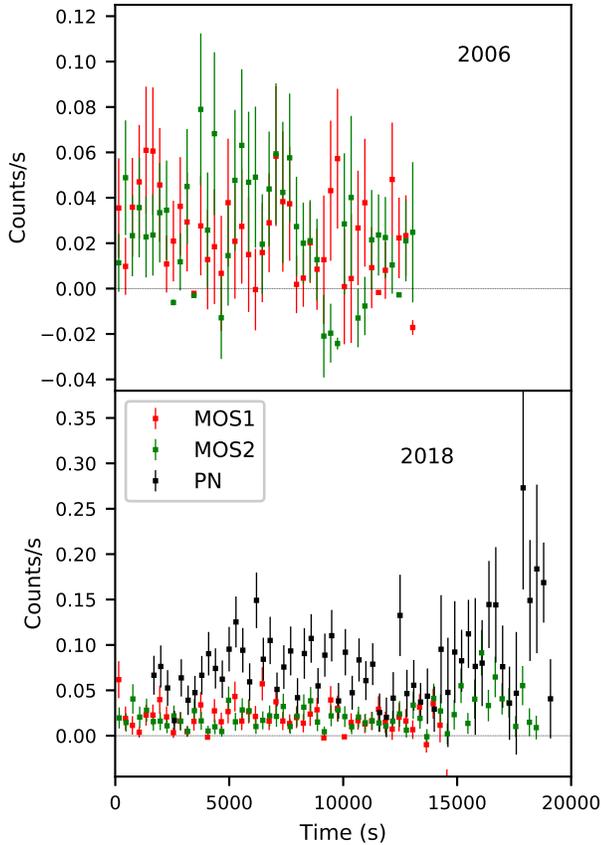}
\caption{X-ray light curves of M311 from 2006 (top) and 2018 (bottom)}
\label{fig:xraylicu}
\end{figure}

In Figure \ref{fig:xraylicu} we show the XMM-Newton X-ray light curves of M311 in all available instruments. There is no obvious evidence of variability.

We reduced the Swift XRT data using \texttt{xrtpipeline} version 0.13.3. The source was not detected in X-rays, and it was outside the field of view of the UV telescope. Using the procedure of \cite{Loredo1992} (eq. 5.13) we obtain a $1\sigma$ upper limit to the count rate of $2.4\times 10^{-4}$\,s$^{-1}$. If we assume the same spectral shape as in the 2018 observation but a different normalisation we obtain a bolometric flux of less than $4.1\times 10^{-13}$ erg\,s$^{-1}$\,cm$^{-2}$. Thus, it seems that M311 is X-ray variable by a factor of at least three.

We also looked for periodicities in the X-ray data for the 2018 observation. To do this, we applied the H-test \citep{deJagerEtAl1989} to the barycenter-corrected EPIC-$pn$ source event list from the 2018 observation. We sought periods between one minute and one hour. We found a significant signal at around 136\,s but, on closer investigation, this turned out to be in the soft proton flaring and not in the source.

We constructed an X-ray to optical ratio by approximating the optical flux by $\log_{10}(F_\text{opt})=-m_\text{V}/2.5-5.37$ \citep{MaccacaroEtAl1988} and comparing this to the X-ray flux between 0.5 and 2.0\,keV. Since M311 is variable both in X-rays and in the optical, this ratio is not well defined. We therefore simply took the 2018 XMM observation, and the approximate non low-state $m_V=16$ magnitude observed by MACHO. We obtained $\log_{10}(F_\text{X}/F_\text{opt})=-2.1$.

\subsection{Optical Monitor}

\begin{figure}
\includegraphics{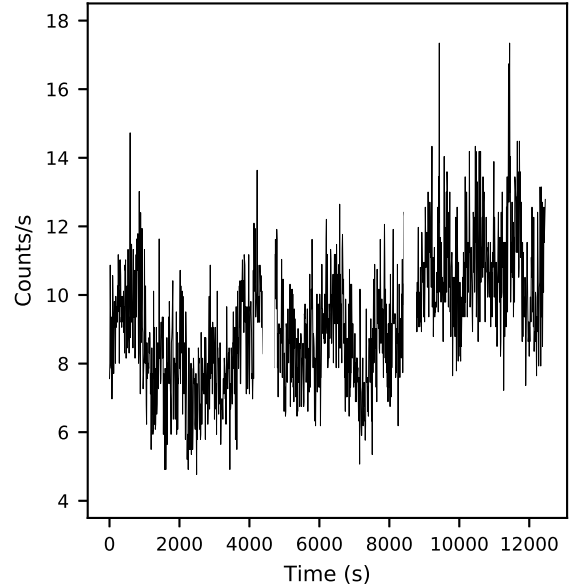}
\caption{OM light curve of M311 from the 2018 XMM-Newton observation, UVW1 filter}
\label{fig:omlicu}
\end{figure}

In the 2018 observation XMM-Newton's Optical Monitor observed the target with the UVW1 filter, centered around 300\,nm. The OM light curve (Figure \ref{fig:omlicu}) showed significant variability over the duration of the observation, ranging from approximately 6 to 12 counts per second. Superimposed on this is apparently some shorter-term flickering of amplitude $\sim 0.1$ mag, but the data is not of sufficient quality to make any definitive statement regarding this flickering. We performed an Analysis-of-Variance (AoV) period search \citep{Schwarzenberg-Czerny1989} but there was no strong signal.  To account for the possibility of longer term variability swamping a fast periodic signal, we subtracted the best-fitting sinusoid from the OM light curve and repeated the AoV search on the remainder. Again, we found no signal.

The magnitude of M311 was $m_{UVW1}=14.8\pm 0.2$, {$M_{UVW1}=4.2\pm 0.3$.} This is quite bright and indicates that the XMM-Newton observation occurred during the high state, and not during one of the dips.

\section{Discussion}

We have studied the optical, UV, and X-ray properties of the X-ray source MACHO 311.37557.169 to attempt to determine its nature. The prominent emission lines of hydrogen suggest that it is a cataclysmic variable, and the long term behaviour of the MACHO light curve resembles a CV switching from high to low states. The decline to the low state, however, would be unusually slow for a magnetic CV. Furthermore, it it was magnetic, we would expect the 4686\AA\ helium line to be stronger. A more likely hypothesis is that it is a VY~Scl star. Its long-term optical light curve strongly resembles, for instance, that of TT~Ari \citep{ZemkoEtAl2014} or V794~Aql \citep{Greiner1998}, both in the depth of the dip and in its duration of a few hundred days. Its high state absolute magnitude of $\sim 5$ is similar to that of V794~Aql, 5.2, and the optical spectrum resembles that of the VY~Scl star RX J2338+431 \citep{WeilEtAl2018}, with the He II 4865 line roughly the same intensity as the He I 4471 line, and showing just a trace of an iron feature at 5169\AA.

The UV light curve from the 2018 observation show features that are consistent with flickering and possibly a superhump period of $\sim 4$\,hours in the UV though the observation is not long enough for a definitive conclusion.

The X-ray spectrum of M311, a partially absorbed two-temperature plasma with a luminosity of order $10^{32}$ to $10^{33}$\,erg\,s$^{-1}$, is also consistent with a CV, which have long been known to be strong X-ray emitters. Again, there are strong similarities to V794~Aql, which also showed a two-temperature plasma spectrum with luminosity $4\times 10^{32}$\,erg\,s$^{-2}$, as calculated from the flux measurement of \cite{ZemkoEtAl2014} together with the distance determination of \cite{Bailer-JonesEtAl2018}. There was some evidence for a fluorescent iron line at 6.4\,keV. The equivalent width was difficult to determine because of the mediocre photon statistics, but seems to be in about the 200 to 400\,eV range. Although this is a higher value than in the VY Scl star TT Ari (0.1\,keV; \citealt{ZemkoEtAl2014}), it appears similar to the one found for V751 Cyg (see \citealt{PageEtAl2014}, Table 2).

If the CV interpretation is correct, then the large absorption column density and covering fraction suggests a high system inclination. There is no evidence for eclipses in any of the light curves, however. The low $F_X/F_\text{opt}$ value of -2.1 is low compared to magnetic CVs, even the X-ray underluminous IPs (e.g., -1.7 in the case of V902~Mon, \citealt{WorpelEtAl2018}).

Other types of variable stars are unlikely. M311 shows Balmer lines, so it is not an R~CrB star. The optical colour is too blue to be a semiregular variable, and the optical reddening with decreasing brightness does not seem to fit a normal dwarf nova or anti dwarf nova. The two-peaked X-ray spectrum superficially resembles that of a $\delta$-type symbiotic variable (e.g., \citealt{LunaEtAl2003}) but M311 is too close for the companion star to be a red giant. Similarly, it is not luminous enough to be a Herbig Ae/Be object. Conversely, it is too X-ray luminous, by at least an order of magnitude, to be a T~Tauri star (e.g. \citealt{TelleschiEtAl2007}, Fig 1). Although the object shows HeII lines their weakness would be unusual for a magnetic CV.

We have found that M311 is likely to be a VY~Scl star, based on its numerous similarities to that class and on ruling out other types of variable stars. A longer optical campaign aimed at determining the orbital, and possibly the spin, periods would be desirable. Further short-term photometry with the goal of finding or ruling out variability on $\sim 15$\,m time scales would also be helpful.

\acknowledgements
This work was supported by the German DLR under contract 50 OR 1814. 
We are grateful to G. Clayton for providing the optical spectra.  
This paper utilizes public domain data obtained by the MACHO Project,
jointly funded by the US Department of Energy through the University
of California, Lawrence Livermore National Laboratory under contract
No. W-7405-Eng-48, by the National Science Foundation through the
Center for Particle Astrophysics of the University of California
under cooperative agreement AST-8809616, and by the Mount Stromlo
and Siding Spring Observatory, part of the Australian National
University. This paper includes data gathered with the 6.5 meter Magellan
Telescopes located at Las Campanas Observatory, Chile.
This research has made use of the SIMBAD database,
operated at CDS, Strasbourg, France. This work has made use of data from the European Space Agency (ESA)
mission {\it Gaia} (\url{https://www.cosmos.esa.int/gaia}), processed by
the {\it Gaia} Data Processing and Analysis Consortium (DPAC,
\url{https://www.cosmos.esa.int/web/gaia/dpac/consortium}). Funding
for the DPAC has been provided by national institutions, in particular
the institutions participating in the {\it Gaia} Multilateral Agreement.
This work has made use of Astropy \citep{Astropy2013, Astropy2018}.
This research has made use of the APASS database, located at the AAVSO web site. Funding for APASS has been provided by the Robert Martin Ayers Sciences Fund.

We are grateful to the referee for suggestions that 
substantially improved the clarity and strength of the paper, for making us aware of the
OGLE data, and for useful advice about the binary system configuration.

\bibliography{bibli}

\begin{thebibliography}{35}
\expandafter\ifx\csname natexlab\endcsname\relax\def\natexlab#1{#1}\fi

\bibitem[{{Alcock} {et~al.}(1997){Alcock}, {Allsman}, {Alves}, {Axelrod},
  {Becker}, {Bennett}, {Cook}, {Freeman}, {Griest}, {Guern}, {Lehner},
  {Marshall}, {Peterson}, {Pratt}, {Quinn}, {Rodgers}, {Stubbs}, {Sutherland},
  \& {Welch}}]{AlcockEtAl1997}
{Alcock}, C., {Allsman}, R.~A., {Alves}, D., {et~al.} 1997, \apj, 486, 697

\bibitem[{{Alcock} {et~al.}(2000){Alcock}, {Allsman}, {Alves}, {Axelrod},
  {Becker}, {Bennett}, {Cook}, {Dalal}, {Drake}, {Freeman}, {Geha}, {Griest},
  {Lehner}, {Marshall}, {Minniti}, {Nelson}, {Peterson}, {Popowski}, {Pratt},
  {Quinn}, {Stubbs}, {Sutherland}, {Tomaney}, {Vandehei}, \&
  {Welch}}]{AlcockEtAl2000}
{Alcock}, C., {Allsman}, R.~A., {Alves}, D.~R., {et~al.} 2000, \apj, 542, 281

\bibitem[{{Alcock} {et~al.}(1992){Alcock}, {Axelrod}, {Bennett}, {Cook},
  {Park}, {Griest}, {Perlmutter}, {Stubbs}, {Freeman}, \&
  {Peterson}}]{AlcockEtAl1992}
{Alcock}, C., {Axelrod}, T.~S., {Bennett}, D.~P., {et~al.} 1992, in
  Astronomical Society of the Pacific Conference Series, Vol.~34, Robotic
  Telescopes in the 1990s, ed. A.~V. {Filippenko}, 193--202

\bibitem[{{Arnaud}(1996)}]{Arnaud1996}
{Arnaud}, K.~A. 1996, in {Astronomical Society of the Pacific Conference
  Series}, Vol. 101, {Astronomical Data Analysis Software and Systems V}, ed.
  {G.~H.~Jacoby \& J.~Barnes}, 17

\bibitem[{{Astropy Collaboration} {et~al.}(2018){Astropy Collaboration},
  {Price-Whelan}, {Sip{\H o}cz}, {G{\"u}nther}, {Lim}, {Crawford}, {Conseil},
  {Shupe}, {Craig}, {Dencheva}, {Ginsburg}, {VanderPlas}, {Bradley},
  {P{\'e}rez-Su{\'a}rez}, {de Val-Borro}, {Aldcroft}, {Cruz}, {Robitaille},
  {Tollerud}, {Ardelean}, {Babej}, {Bach}, {Bachetti}, {Bakanov}, {Bamford},
  {Barentsen}, {Barmby}, {Baumbach}, {Berry}, {Biscani}, {Boquien}, {Bostroem},
  {Bouma}, {Brammer}, {Bray}, {Breytenbach}, {Buddelmeijer}, {Burke},
  {Calderone}, {Cano Rodr{\'{\i}}guez}, {Cara}, {Cardoso}, {Cheedella},
  {Copin}, {Corrales}, {Crichton}, {D'Avella}, {Deil}, {Depagne}, {Dietrich},
  {Donath}, {Droettboom}, {Earl}, {Erben}, {Fabbro}, {Ferreira}, {Finethy},
  {Fox}, {Garrison}, {Gibbons}, {Goldstein}, {Gommers}, {Greco}, {Greenfield},
  {Groener}, {Grollier}, {Hagen}, {Hirst}, {Homeier}, {Horton}, {Hosseinzadeh},
  {Hu}, {Hunkeler}, {Ivezi{\'c}}, {Jain}, {Jenness}, {Kanarek}, {Kendrew},
  {Kern}, {Kerzendorf}, {Khvalko}, {King}, {Kirkby}, {Kulkarni}, {Kumar},
  {Lee}, {Lenz}, {Littlefair}, {Ma}, {Macleod}, {Mastropietro}, {McCully},
  {Montagnac}, {Morris}, {Mueller}, {Mumford}, {Muna}, {Murphy}, {Nelson},
  {Nguyen}, {Ninan}, {N{\"o}the}, {Ogaz}, {Oh}, {Parejko}, {Parley}, {Pascual},
  {Patil}, {Patil}, {Plunkett}, {Prochaska}, {Rastogi}, {Reddy Janga},
  {Sabater}, {Sakurikar}, {Seifert}, {Sherbert}, {Sherwood-Taylor}, {Shih},
  {Sick}, {Silbiger}, {Singanamalla}, {Singer}, {Sladen}, {Sooley},
  {Sornarajah}, {Streicher}, {Teuben}, {Thomas}, {Tremblay}, {Turner},
  {Terr{\'o}n}, {van Kerkwijk}, {de la Vega}, {Watkins}, {Weaver}, {Whitmore},
  {Woillez}, {Zabalza}, \& {Astropy Contributors}}]{Astropy2018}
{Astropy Collaboration}, {Price-Whelan}, A.~M., {Sip{\H o}cz}, B.~M., {et~al.}
  2018, \aj, 156, 123

\bibitem[{{Astropy Collaboration} {et~al.}(2013){Astropy Collaboration},
  {Robitaille}, {Tollerud}, {Greenfield}, {Droettboom}, {Bray}, {Aldcroft},
  {Davis}, {Ginsburg}, {Price-Whelan}, {Kerzendorf}, {Conley}, {Crighton},
  {Barbary}, {Muna}, {Ferguson}, {Grollier}, {Parikh}, {Nair}, {Unther},
  {Deil}, {Woillez}, {Conseil}, {Kramer}, {Turner}, {Singer}, {Fox}, {Weaver},
  {Zabalza}, {Edwards}, {Azalee Bostroem}, {Burke}, {Casey}, {Crawford},
  {Dencheva}, {Ely}, {Jenness}, {Labrie}, {Lim}, {Pierfederici}, {Pontzen},
  {Ptak}, {Refsdal}, {Servillat}, \& {Streicher}}]{Astropy2013}
{Astropy Collaboration}, {Robitaille}, T.~P., {Tollerud}, E.~J., {et~al.} 2013,
  \aap, 558, A33

\bibitem[{{Bailer-Jones} {et~al.}(2018){Bailer-Jones}, {Rybizki}, {Fouesneau},
  {Mantelet}, \& {Andrae}}]{Bailer-JonesEtAl2018}
{Bailer-Jones}, C.~A.~L., {Rybizki}, J., {Fouesneau}, M., {Mantelet}, G., \&
  {Andrae}, R. 2018, \aj, 156, 58

\bibitem[{{Cieslinski} {et~al.}(2004){Cieslinski}, {Diaz}, {Drake}, \&
  {Cook}}]{CieslinskiEtAl2004}
{Cieslinski}, D., {Diaz}, M.~P., {Drake}, A.~J., \& {Cook}, K.~H. 2004, \pasp,
  116, 610

\bibitem[{{Collins} {et~al.}(2017){Collins}, {Kielkopf}, {Stassun}, \&
  {Hessman}}]{CollinsEtAl2017}
{Collins}, K.~A., {Kielkopf}, J.~F., {Stassun}, K.~G., \& {Hessman}, F.~V.
  2017, \aj, 153, 77

\bibitem[{{de Jager} {et~al.}(1989){de Jager}, {Raubenheimer}, \&
  {Swanepoel}}]{deJagerEtAl1989}
{de Jager}, O.~C., {Raubenheimer}, B.~C., \& {Swanepoel}, J.~W.~H. 1989, \aap,
  221, 180

\bibitem[{{Eastman} {et~al.}(2010){Eastman}, {Siverd}, \&
  {Gaudi}}]{EastmanEtAl2010}
{Eastman}, J., {Siverd}, R., \& {Gaudi}, B.~S. 2010, \pasp, 122, 935

\bibitem[{{Gaia Collaboration} {et~al.}(2018){Gaia Collaboration}, {Brown},
  {Vallenari}, {Prusti}, {de Bruijne}, {Babusiaux}, \&
  {Bailer-Jones}}]{Gaia2018}
{Gaia Collaboration}, {Brown}, A.~G.~A., {Vallenari}, A., {et~al.} 2018, ArXiv
  e-prints

\bibitem[{{Gaia Collaboration} {et~al.}(2016){Gaia Collaboration}, {Prusti},
  {de Bruijne}, {Brown}, {Vallenari}, {Babusiaux}, {Bailer-Jones}, {Bastian},
  {Biermann}, {Evans}, {Eyer}, {Jansen}, {Jordi}, {Klioner}, {Lammers},
  {Lindegren}, {Luri}, {Mignard}, {Milligan}, {Panem}, {Poinsignon},
  {Pourbaix}, {Randich}, {Sarri}, {Sartoretti}, {Siddiqui}, {Soubiran},
  {Valette}, {van Leeuwen}, {Walton}, {Aerts}, {Arenou}, {Cropper}, {Drimmel},
  {H{\o}g}, {Katz}, {Lattanzi}, {O'Mullane}, {Grebel}, {Holland}, {Huc},
  {Passot}, {Bramante}, {Cacciari}, {Casta{\~n}eda}, {Chaoul}, {Cheek}, {De
  Angeli}, {Fabricius}, {Guerra}, {Hern{\'a}ndez}, {Jean-Antoine-Piccolo},
  {Masana}, {Messineo}, {Mowlavi}, {Nienartowicz}, {Ord{\'o}{\~n}ez- Blanco},
  {Panuzzo}, {Portell}, {Richards}, {Riello}, {Seabroke}, {Tanga},
  {Th{\'e}venin}, {Torra}, {Els}, {Gracia- Abril}, {Comoretto},
  {Garcia-Reinaldos}, {Lock}, {Mercier}, {Altmann}, {Andrae}, {Astraatmadja},
  {Bellas-Velidis}, {Benson}, {Berthier}, {Blomme}, {Busso}, {Carry},
  {Cellino}, {Clementini}, {Cowell}, {Creevey}, {Cuypers}, {Davidson}, {De
  Ridder}, {de Torres}, {Delchambre}, {Dell'Oro}, {Ducourant}, {Fr{\'e}mat},
  {Garc{\'\i}a-Torres}, {Gosset}, {Halbwachs}, {Hambly}, {Harrison}, {Hauser},
  {Hestroffer}, {Hodgkin}, {Huckle}, {Hutton}, {Jasniewicz}, {Jordan},
  {Kontizas}, {Korn}, {Lanzafame}, {Manteiga}, {Moitinho}, {Muinonen},
  {Osinde}, {Pancino}, {Pauwels}, {Petit}, {Recio-Blanco}, {Robin}, {Sarro},
  {Siopis}, {Smith}, {Smith}, {Sozzetti}, {Thuillot}, {van Reeven}, {Viala},
  {Abbas}, {Abreu Aramburu}, {Accart}, {Aguado}, {Allan}, {Allasia},
  {Altavilla}, {{\'A}lvarez}, {Alves}, {Anderson}, {Andrei}, {Anglada Varela},
  {Antiche}, {Antoja}, {Ant{\'o}n}, {Arcay}, {Atzei}, {Ayache}, {Bach},
  {Baker}, {Balaguer-N{\'u}{\~n}ez}, {Barache}, {Barata}, {Barbier}, {Barblan},
  {Baroni}, {Barrado y Navascu{\'e}s}, {Barros}, {Barstow}, {Becciani},
  {Bellazzini}, {Bellei}, {Bello Garc{\'\i}a}, {Belokurov}, {Bendjoya},
  {Berihuete}, {Bianchi}, {Bienaym{\'e}}, {Billebaud}, {Blagorodnova},
  {Blanco-Cuaresma}, {Boch}, {Bombrun}, {Borrachero}, {Bouquillon}, {Bourda},
  {Bouy}, {Bragaglia}, {Breddels}, {Brouillet}, {Br{\"u}semeister},
  {Bucciarelli}, {Budnik}, {Burgess}, {Burgon}, {Burlacu}, {Busonero}, {Buzzi},
  {Caffau}, {Cambras}, {Campbell}, {Cancelliere}, {Cantat-Gaudin}, {Carlucci},
  {Carrasco}, {Castellani}, {Charlot}, {Charnas}, {Charvet}, {Chassat},
  {Chiavassa}, {Clotet}, {Cocozza}, {Collins}, {Collins}, {Costigan}, {Crifo},
  {Cross}, {Crosta}, {Crowley}, {Dafonte}, {Damerdji}, {Dapergolas}, {David},
  {David}, {De Cat}, {de Felice}, {de Laverny}, {De Luise}, {De March}, {de
  Martino}, {de Souza}, {Debosscher}, {del Pozo}, {Delbo}, {Delgado},
  {Delgado}, {di Marco}, {Di Matteo}, {Diakite}, {Distefano}, {Dolding}, {Dos
  Anjos}, {Drazinos}, {Dur{\'a}n}, {Dzigan}, {Ecale}, {Edvardsson}, {Enke},
  {Erdmann}, {Escolar}, {Espina}, {Evans}, {Eynard Bontemps}, {Fabre},
  {Fabrizio}, {Faigler}, {Falc{\~a}o}, {Farr{\`a}s Casas}, {Faye}, {Federici},
  {Fedorets}, {Fern{\'a}ndez-Hern{\'a}ndez}, {Fernique}, {Fienga}, {Figueras},
  {Filippi}, {Findeisen}, {Fonti}, {Fouesneau}, {Fraile}, {Fraser}, {Fuchs},
  {Furnell}, {Gai}, {Galleti}, {Galluccio}, {Garabato}, {Garc{\'\i}a-Sedano},
  {Gar{\'e}}, {Garofalo}, {Garralda}, {Gavras}, {Gerssen}, {Geyer}, {Gilmore},
  {Girona}, {Giuffrida}, {Gomes}, {Gonz{\'a}lez-Marcos},
  {Gonz{\'a}lez-N{\'u}{\~n}ez}, {Gonz{\'a}lez-Vidal}, {Granvik}, {Guerrier},
  {Guillout}, {Guiraud}, {G{\'u}rpide}, {Guti{\'e}rrez-S{\'a}nchez}, {Guy},
  {Haigron}, {Hatzidimitriou}, {Haywood}, {Heiter}, {Helmi}, {Hobbs},
  {Hofmann}, {Holl}, {Holland}, {Hunt}, {Hypki}, {Icardi}, {Irwin}, {Jevardat
  de Fombelle}, {Jofr{\'e}}, {Jonker}, {Jorissen}, {Julbe}, {Karampelas},
  {Kochoska}, {Kohley}, {Kolenberg}, {Kontizas}, {Koposov}, {Kordopatis},
  {Koubsky}, {Kowalczyk}, {Krone-Martins}, {Kudryashova}, {Kull}, {Bachchan},
  {Lacoste-Seris}, {Lanza}, {Lavigne}, {Le Poncin-Lafitte}, {Lebreton},
  {Lebzelter}, {Leccia}, {Leclerc}, {Lecoeur-Taibi}, {Lemaitre}, {Lenhardt},
  {Leroux}, {Liao}, {Licata}, {Lindstr{\o}m}, {Lister}, {Livanou}, {Lobel},
  {L{\"o}ffler}, {L{\'o}pez}, {Lopez-Lozano}, {Lorenz}, {Loureiro},
  {MacDonald}, {Magalh{\~a}es Fernandes}, {Managau}, {Mann}, {Mantelet},
  {Marchal}, {Marchant}, {Marconi}, {Marie}, {Marinoni}, {Marrese},
  {Marschalk{\'o}}, {Marshall}, {Mart{\'\i}n-Fleitas}, {Martino}, {Mary},
  {Matijevi{\v{c}}}, {Mazeh}, {McMillan}, {Messina}, {Mestre}, {Michalik},
  {Millar}, {Miranda}, {Molina}, {Molinaro}, {Molinaro}, {Moln{\'a}r},
  {Moniez}, {Montegriffo}, {Monteiro}, {Mor}, {Mora}, {Morbidelli}, {Morel},
  {Morgenthaler}, {Morley}, {Morris}, {Mulone}, {Muraveva}, {Musella},
  {Narbonne}, {Nelemans}, {Nicastro}, {Noval}, {Ord{\'e}novic},
  {Ordieres-Mer{\'e}}, {Osborne}, {Pagani}, {Pagano}, {Pailler}, {Palacin},
  {Palaversa}, {Parsons}, {Paulsen}, {Pecoraro}, {Pedrosa}, {Pentik{\"a}inen},
  {Pereira}, {Pichon}, {Piersimoni}, {Pineau}, {Plachy}, {Plum}, {Poujoulet},
  {Pr{\v{s}}a}, {Pulone}, {Ragaini}, {Rago}, {Rambaux}, {Ramos-Lerate},
  {Ranalli}, {Rauw}, {Read}, {Regibo}, {Renk}, {Reyl{\'e}}, {Ribeiro},
  {Rimoldini}, {Ripepi}, {Riva}, {Rixon}, {Roelens}, {Romero-G{\'o}mez},
  {Rowell}, {Royer}, {Rudolph}, {Ruiz-Dern}, {Sadowski}, {Sagrist{\`a}
  Sell{\'e}s}, {Sahlmann}, {Salgado}, {Salguero}, {Sarasso}, {Savietto},
  {Schnorhk}, {Schultheis}, {Sciacca}, {Segol}, {Segovia}, {Segransan},
  {Serpell}, {Shih}, {Smareglia}, {Smart}, {Smith}, {Solano}, {Solitro},
  {Sordo}, {Soria Nieto}, {Souchay}, {Spagna}, {Spoto}, {Stampa}, {Steele},
  {Steidelm{\"u}ller}, {Stephenson}, {Stoev}, {Suess}, {S{\"u}veges}, {Surdej},
  {Szabados}, {Szegedi-Elek}, {Tapiador}, {Taris}, {Tauran}, {Taylor},
  {Teixeira}, {Terrett}, {Tingley}, {Trager}, {Turon}, {Ulla}, {Utrilla},
  {Valentini}, {van Elteren}, {Van Hemelryck}, {van Leeuwen}, {Varadi},
  {Vecchiato}, {Veljanoski}, {Via}, {Vicente}, {Vogt}, {Voss}, {Votruba},
  {Voutsinas}, {Walmsley}, {Weiler}, {Weingrill}, {Werner}, {Wevers},
  {Whitehead}, {Wyrzykowski}, {Yoldas}, {{\v{Z}}erjal}, {Zucker}, {Zurbach},
  {Zwitter}, {Alecu}, {Allen}, {Allende Prieto}, {Amorim},
  {Anglada-Escud{\'e}}, {Arsenijevic}, {Azaz}, {Balm}, {Beck}, {Bernstein},
  {Bigot}, {Bijaoui}, {Blasco}, {Bonfigli}, {Bono}, {Boudreault}, {Bressan},
  {Brown}, {Brunet}, {Bunclark}, {Buonanno}, {Butkevich}, {Carret}, {Carrion},
  {Chemin}, {Ch{\'e}reau}, {Corcione}, {Darmigny}, {de Boer}, {de Teodoro}, {de
  Zeeuw}, {Delle Luche}, {Domingues}, {Dubath}, {Fodor}, {Fr{\'e}zouls},
  {Fries}, {Fustes}, {Fyfe}, {Gallardo}, {Gallegos}, {Gardiol}, {Gebran},
  {Gomboc}, {G{\'o}mez}, {Grux}, {Gueguen}, {Heyrovsky}, {Hoar}, {Iannicola},
  {Isasi Parache}, {Janotto}, {Joliet}, {Jonckheere}, {Keil}, {Kim},
  {Klagyivik}, {Klar}, {Knude}, {Kochukhov}, {Kolka}, {Kos}, {Kutka}, {Lainey},
  {LeBouquin}, {Liu}, {Loreggia}, {Makarov}, {Marseille}, {Martayan},
  {Martinez-Rubi}, {Massart}, {Meynadier}, {Mignot}, {Munari}, {Nguyen},
  {Nordlander}, {Ocvirk}, {O'Flaherty}, {Olias Sanz}, {Ortiz}, {Osorio},
  {Oszkiewicz}, {Ouzounis}, {Palmer}, {Park}, {Pasquato}, {Peltzer}, {Peralta},
  {P{\'e}turaud}, {Pieniluoma}, {Pigozzi}, {Poels}, {Prat}, {Prod'homme},
  {Raison}, {Rebordao}, {Risquez}, {Rocca-Volmerange}, {Rosen}, {Ruiz-Fuertes},
  {Russo}, {Sembay}, {Serraller Vizcaino}, {Short}, {Siebert}, {Silva},
  {Sinachopoulos}, {Slezak}, {Soffel}, {Sosnowska}, {Strai{\v{z}}ys}, {ter
  Linden}, {Terrell}, {Theil}, {Tiede}, {Troisi}, {Tsalmantza}, {Tur},
  {Vaccari}, {Vachier}, {Valles}, {Van Hamme}, {Veltz}, {Virtanen}, {Wallut},
  {Wichmann}, {Wilkinson}, {Ziaeepour}, \& {Zschocke}}]{Gaia2016}
{Gaia Collaboration}, {Prusti}, T., {de Bruijne}, J.~H.~J., {et~al.} 2016,
  \aap, 595

\bibitem[{{Greiner}(1998)}]{Greiner1998}
{Greiner}, J. 1998, \aap, 336, 626

\bibitem[{{Henden} {et~al.}(2018){Henden}, {Levine}, {Terrell}, {Welch},
  {Munari}, \& {Kloppenborg}}]{HendenEtAl2018}
{Henden}, A.~A., {Levine}, S., {Terrell}, D., {et~al.} 2018, in American
  Astronomical Society Meeting Abstracts, Vol. 232, American Astronomical
  Society Meeting Abstracts \#232, 223.06

\bibitem[{{Hoffleit}(1972)}]{Hoffleit1972}
{Hoffleit}, D. 1972, Information Bulletin on Variable Stars, 660

\bibitem[{{Kaastra}(2017)}]{Kaastra2017}
{Kaastra}, J.~S. 2017, \aap, 605, A51

\bibitem[{{Kochanek} {et~al.}(2017){Kochanek}, {Shappee}, {Stanek}, {Holoien},
  {Thompson}, {Prieto}, {Dong}, {Shields}, {Will}, {Britt}, {Perzanowski}, \&
  {Pojma{\'n}ski}}]{KochanekEtAl2017}
{Kochanek}, C.~S., {Shappee}, B.~J., {Stanek}, K.~Z., {et~al.} 2017, \pasp,
  129, 104502

\bibitem[{{Liedahl} {et~al.}(1995){Liedahl}, {Osterheld}, \&
  {Goldstein}}]{LiedahlEtAl1995}
{Liedahl}, D.~A., {Osterheld}, A.~L., \& {Goldstein}, W.~H. 1995, \apjl, 438,
  L115

\bibitem[{{Loredo}(1992)}]{Loredo1992}
{Loredo}, T.~J. 1992, in Statistical Challenges in Modern Astronomy, ed. E.~D.
  {Feigelson} \& G.~J. {Babu}, 275--297

\bibitem[{{Luna} {et~al.}(2013){Luna}, {Sokoloski}, {Mukai}, \&
  {Nelson}}]{LunaEtAl2003}
{Luna}, G.~J.~M., {Sokoloski}, J.~L., {Mukai}, K., \& {Nelson}, T. 2013, \aap,
  559, A6

\bibitem[{{Maccacaro} {et~al.}(1988){Maccacaro}, {Gioia}, {Wolter}, {Zamorani},
  \& {Stocke}}]{MaccacaroEtAl1988}
{Maccacaro}, T., {Gioia}, I.~M., {Wolter}, A., {Zamorani}, G., \& {Stocke},
  J.~T. 1988, \apj, 326, 680

\bibitem[{{Mewe} {et~al.}(1985){Mewe}, {Gronenschild}, \& {van den
  Oord}}]{MeweEtAl1985}
{Mewe}, R., {Gronenschild}, E.~H.~B.~M., \& {van den Oord}, G.~H.~J. 1985,
  \aaps, 62, 197

\bibitem[{{Page} {et~al.}(2014){Page}, {Osborne}, {Beardmore}, {Evans},
  {Rosen}, \& {Watson}}]{PageEtAl2014}
{Page}, K.~L., {Osborne}, J.~P., {Beardmore}, A.~P., {et~al.} 2014, \aap, 570,
  A37

\bibitem[{{Popowski} {et~al.}(2003){Popowski}, {Cook}, \&
  {Becker}}]{PopowskiEtAl2003}
{Popowski}, P., {Cook}, K.~H., \& {Becker}, A.~C. 2003, \aj, 126, 2910

\bibitem[{{Rosen} {et~al.}(2016){Rosen}, {Webb}, {Watson}, {Ballet}, {Barret},
  {Braito}, {Carrera}, {Ceballos}, {Coriat}, {Della Ceca}, {Denkinson},
  {Esquej}, {Farrell}, {Freyberg}, {Gris{\'e}}, {Guillout}, {Heil},
  {Koliopanos}, {Law-Green}, {Lamer}, {Lin}, {Martino}, {Michel}, {Motch},
  {Nebot Gomez-Moran}, {Page}, {Page}, {Page}, {Pakull}, {Pye}, {Read},
  {Rodriguez}, {Sakano}, {Saxton}, {Schwope}, {Scott}, {Sturm}, {Traulsen},
  {Yershov}, \& {Zolotukhin}}]{RosenEtAl2016}
{Rosen}, S.~R., {Webb}, N.~A., {Watson}, M.~G., {et~al.} 2016, \aap, 590, A1

\bibitem[{{Samus} {et~al.}(2004){Samus}, {Durlevich}, \& {et
  al.}}]{SamusEtAl2004}
{Samus}, N.~N., {Durlevich}, O.~V., \& {et al.} 2004, VizieR Online Data
  Catalog

\bibitem[{{Schwarzenberg-Czerny}(1989)}]{Schwarzenberg-Czerny1989}
{Schwarzenberg-Czerny}, A. 1989, \mnras, 241, 153

\bibitem[{{Shappee} {et~al.}(2014){Shappee}, {Prieto}, {Grupe}, {Kochanek},
  {Stanek}, {De Rosa}, {Mathur}, {Zu}, {Peterson}, {Pogge}, {Komossa}, {Im},
  {Jencson}, {Holoien}, {Basu}, {Beacom}, {Szczygie{\l}}, {Brimacombe},
  {Adams}, {Campillay}, {Choi}, {Contreras}, {Dietrich}, {Dubberley},
  {Elphick}, {Foale}, {Giustini}, {Gonzalez}, {Hawkins}, {Howell}, {Hsiao},
  {Koss}, {Leighly}, {Morrell}, {Mudd}, {Mullins}, {Nugent}, {Parrent},
  {Phillips}, {Pojmanski}, {Rosing}, {Ross}, {Sand}, {Terndrup}, {Valenti},
  {Walker}, \& {Yoon}}]{ShappeeEtAL2014}
{Shappee}, B.~J., {Prieto}, J.~L., {Grupe}, D., {et~al.} 2014, \apj, 788, 48

\bibitem[{{Szyma{\'n}ski} {et~al.}(2011){Szyma{\'n}ski}, {Udalski},
  {Soszy{\'n}ski}, {Kubiak}, {Pietrzy{\'n}ski}, {Poleski}, {Wyrzykowski}, \&
  {Ulaczyk}}]{SzymanskiEtAl2011}
{Szyma{\'n}ski}, M.~K., {Udalski}, A., {Soszy{\'n}ski}, I., {et~al.} 2011,
  \actaa, 61, 83

\bibitem[{{Telleschi} {et~al.}(2007){Telleschi}, {G{\"u}del}, {Briggs},
  {Audard}, \& {Palla}}]{TelleschiEtAl2007}
{Telleschi}, A., {G{\"u}del}, M., {Briggs}, K.~R., {Audard}, M., \& {Palla}, F.
  2007, \aap, 468, 425

\bibitem[{{Weil} {et~al.}(2018){Weil}, {Thorstensen}, \&
  {Haberl}}]{WeilEtAl2018}
{Weil}, K.~E., {Thorstensen}, J.~R., \& {Haberl}, F. 2018, ArXiv e-prints

\bibitem[{{Worpel} {et~al.}(2018){Worpel}, {Schwope}, {Traulsen}, {Mukai}, \&
  {Ok}}]{WorpelEtAl2018}
{Worpel}, H., {Schwope}, A.~D., {Traulsen}, I., {Mukai}, K., \& {Ok}, S. 2018,
  \aap, 617, A52

\bibitem[{{Zaniewski} {et~al.}(2005){Zaniewski}, {Clayton}, {Welch}, {Gordon},
  {Minniti}, \& {Cook}}]{ZaniewskiEtAl2005}
{Zaniewski}, A., {Clayton}, G.~C., {Welch}, D.~L., {et~al.} 2005, \aj, 130,
  2293

\bibitem[{{Zemko} {et~al.}(2014){Zemko}, {Orio}, {Mukai}, \&
  {Shugarov}}]{ZemkoEtAl2014}
{Zemko}, P., {Orio}, M., {Mukai}, K., \& {Shugarov}, S. 2014, \mnras, 445, 869

\end{thebibliography}

\end{document}